# How much of the Solar System should we leave as Wilderness?

Martin Elvis[a][1] and Tony Milligan[b]

a.  Center for Astrophysics | Harvard & Smithsonian, 60 Garden St.,
    Cambridge MA 02138, USA; melvis@cfa.harvard.edu;
b.  Department of Theology and Religious Studies, King's College London,
    Virginia Woolf Building, 22 Kingsway, London WC2B 6NR;
    anthony.milligan@kcl.ac.uk

## ABSTRACT

*"How much of the Solar System should we reserve as wilderness, off-limits to human development?" We make a general argument that, as a matter of fixed policy, development should be limited to one eighth, with the remainder set aside. We argue that adopting a "one-eighth principle" is far less restrictive, overall, than it might seem. One eighth of the iron in the asteroid belt is more than a million times greater than all of the Earth's currently estimated iron ore reserves, and it may well suffice for centuries. A limit of some sort is necessary because of the problems associated with exponential growth. We note that humans are poor at estimating the pace of such growth and, as a result, the limitations of a resource are hard to recognize before the final three doubling times. These three doublings take utilization successively from an eighth to a quarter, then to a half, and then to the point of exhaustion. Population growth and climate change are instances of unchecked exponential growth. Each places strains upon our available resources, each is a recognized problem that we would like to control, but attempts to do so at this comparatively late stage in their development have not been encouraging.*

*Our limited ability to see ahead until such processes are far advanced, suggests that we should set ourselves a "tripwire" that gives us at least 3 doubling times as leeway. This tripwire would be triggered when one eighth of the Solar System's resources are close to being exploited. The timescale on which we might hit this tripwire, for several assumed growth rates, is long. At a 3.5% growth rate for the space economy, comparable to that of the use of iron from the beginning of the Industrial Revolution until now, the $1/8^{th}$ point would be reached after 400 years. At that point, the 20-year doubling time associated with a 3.5% growth rate would mean that only 60 years would remain in which to transition the economic system to new "steady state" conditions. The rationale for adopting the one-eighth principle so far in advance is that it may be far easier to implement in-principle restrictions at an early stage, rather than later, when vested and competing interests have come into existence under conditions of diminishing opportunity.*

---

[1] Corresponding author.

## 1. INTRODUCTION

The Solar System is big[2]. It is so big that the idea that humans may fully exploit and deplete its resources seems absurd. Yet if a true economy emerges in space it will start to make use of the vast yet finite resources of the Moon, Mars and small Solar System bodies (such as asteroids)[3]. We have no good reason to believe that such an off-world economy would behave in a radically different way from terrestrial economies and the latter (as we know) grow exponentially. After a century of reasonably modest 3.5% annual growth, any economy will be nearly 20 times larger than it was to begin with. If the off-world economy proves to be especially dynamic, and grows at more than 3.5%, this may be commercially advantageous in the short term, but could lead to problems of resource depletion or exhaustion surprisingly soon. Once we have exploited our solar system, there is no other plausible and accessible new frontier[4]. In what follows, we will refer to the point where untapped resources cannot readily be brought into use, as the point of "super-exploitation."

Such a state of affairs would risk a crisis of potentially catastrophic proportions as any industries reliant upon the incorporation of new resources would simply run dry. It is tempting to regard the danger as one of a Malthusian crisis in space, i.e. a crisis which emerges out of growth, followed by a steady push towards successively more marginal resources, followed by exhaustion. We will not, however, quibble about the classification. Approaching a point of super-exploitation is something that we *ought* to be concerned about if we assume that we ought to be concerned, *at this point in time and in action-guiding ways*, not only about ourselves but about future generations of humans (e.g. [1,2]). More precisely, humans whose lives we can influence in ways that are (up to a point) predictably advantageous or disadvantageous. This will apply *at least* to those who live within the next 500 years. Beyond some limit, the unpredictable long-term impact of our actions may make it difficult to include the interests of future humans within our deliberations. However, those who appeal to the future of humanity as a justification for space exploration (or, indeed for any action whatsoever) should accept *at least* concern for future humans within this limited time-scale. In what follows, we are committed to accepting it.

As a way of avoiding a point of super-exploitation, we will present a case for adopting a precautionary "one-eighth principle" with regard to the exploitable materials of the Solar System and, more specifically, with regard to its solid bodies. As a provisional statement of the principle, the following captures the key intent:

---

[2] As classically stated by Douglas Adams: "Space is big. You just won't believe how vastly, hugely, mind- bogglingly big it is. I mean, you may think it's a long way down the road to the chemist's, but that's just peanuts to space", *The Hitchhiker's Guide to the Galaxy.* Numerically, the distance from Earth to Mars at their closest approach [3] is 1350 times longer than a round-the-world voyage (~40,000 km). Compared to the outer planets, Mars is a near neighbor.

[3] [4,5] begin to grapple with finitude of Solar System resources.

[4] Unless new physics, now unimagined, allows rapid transit to and from other planetary systems.

**The one-eighth principle**: *While economic growth remains exponential, we should regard as ours to use no more than one-eighth of the exploitable materials of the Solar System. And by 'ours' we mean humanity's as a whole, rather than any particular generation of humans or group of generations. The remaining seven-eighths of the exploitable Solar System should be left as space wilderness.*

The growth rates we refer to are in the use of fresh resources, such as iron. Any recycling of resources will reduce these growth rates and stretch out the timescales. It is clear, in fact, from the analysis in this paper that such efforts should be 'baked into' our expansion into space. This is likely to be the case as truly "circular economies" [6], where everything is recycled except sunlight, may well be realized first for in-space habitats that have to be independent of Earth for years at a time, for example to send humans to Mars[5].

The motivation for this principle is the subject of this paper. The restriction of the principle to 'exploitable materials' excludes the constituent materials of the Sun, while they remain constitutive and are not simply emissions. Similarly, if it turns out that the mass of Jupiter (which is greater than that of all the other planets combined) would generate insurmountable problems for ourselves or any future humans to use it as a resource, because of the energy required to escape from its gravity well, then the constitutive materials of Jupiter might also have to be regarded as excluded. What we have in mind, instead, are the *exploitable* planets and their atmospheres, moons, and rings, plus the comets and asteroids. We claim that we should use, at most, one-eighth of these.

As a further qualification, if growth is *not* exponential, i.e. if we ever reach a stable-state economic system, without any danger of collapsing back into exponential growth, or if we develop some effective and *reliable* overall breaking-mechanism which would allow us to transition at any preferred time from exponential growth to a stable state system then the one-eighth principle might reasonably be set aside.

Otherwise, we will argue that this principle sets a suitably prudent maximum limit. The one-eighth principle does not, however, underpin any broader argument against economic development and growth. A more restrictive 1/16 or 1/32 principle has the problem that a minor error in estimating the growth rate can lead to a major error in predicting when super-exploitation will be reached. Ours is not an anti-growth argument, and it is consistent with different views about how much of the Earth's resources ought to be brought into use. It excludes only unconstrained or runaway growth. The principle would, in fact, be redundant if there was some broader case against all economic growth.

---

[5] E.g. NASA closed-loop environmental control and life support systems (ECLSS): https://www.nasa.gov/sites/default/files/atoms/files/life_support_systems.pdf

On the other hand, the principle does not on its own support an automatic entitlement to actually use all one-eighth of the materials of any particular object, e.g. the Moon or Mars. There may well be locations that require stronger protection because of, e.g. their uniqueness, as we discuss below. The one-eighth principle simply establishes a prudent *upper* limit for the legitimacy of growth. How close to this limit it is actually wise to go, overall or in any particular location, will depend upon a broader set of economic, social and ethical considerations. For simplicity, we will also treat the Solar System as a closed system with only negligible movement *in* and *out*. The difficulties of interstellar travel make this a good assumption for the foreseeable future[6], notwithstanding the recent detection of an interstellar asteroid passing through the solar system (7).

## 2. WILDERNESS

As formulated above, the one-eighth principle refers specifically to "wilderness" rather than, for example, "unused materials", "territory", or "pristine environments." In doing so it presupposes a level of continuity between constraint in space and established forms of environmental protection, in particular the path-breaking 1964 US Wilderness Act [8]. There have, however, already been several attempts to apply the wilderness concept to space [9,10] employing various different conceptions of what a concept of wilderness is *for*, i.e. the roles that it is expected to play. As our primary concern here is the avoidance of resource depletion rather than the protection of the natural against human activity, we will draw only upon a "thin" concept of wilderness that excludes various sorts of human *use* but not all forms of human *impact.* Space may, in fact, be the only place where a more demanding conception of wilderness as "pristine environment" is now viable. Everywhere on Earth has become a poor fit. However, as we must either deny that there is any wilderness left on Earth, or else accept that wilderness is compatible with at least some level of human impact[7] (e.g. changes in the levels of $CO_2$ in the atmosphere and absorption of the latter by rocks) it will make sense to adopt a unifying thin wilderness concept.

There are further, obvious, advantages to this move. By the end of the present century, both the lunar and Martian surfaces *as a whole* are likely to be somewhat affected by our presence. The Martian wind (even without exaggeration of its scale) will ensure that any local contamination on a single area of Mars will eventually have at least some global impact upon the planetary surface. Near-pristine lunar parks could be easier to sustain for a period of time, given the different local conditions on the Moon, but they too might eventually be affected *to a degree* by a sustained human presence. On the proposed approach, both would still remain good candidates for wilderness protection. The form taken by such protection is a rather different matter. Areas of Mars might, for example and as suggested by [10,11], be set aside as designated "Planetary Parks," drawing upon the model of National Parks in countries such as the U.S., Australia, Canada and Russia. As in such terrestrial areas, some level of human influence may be presumed. For both Mars and the Moon, the impact in question need not change the standing of protected areas.

---

[6] I.e. excluding some new physics breakthrough.
[7] As a clarification, the thin concept of wilderness that we are using is consistent with certain kinds of prior human impact, but it does not require prior human impact. This sets it at a distance from any thicker concept of wilderness as *what remains once indigenous peoples have been cleared from the land.*

In line with the 1964 Wilderness Act, we will regard wilderness as areas with *limited* human impact. Such places are areas where "man himself is a visitor who does not remain" (U.S. Congress, 1964, sec. 2(c))[8]. They may still be studied *in situ* (with qualifications added about visitor status) and various kinds of "wilderness experience" remain possible just so long as they are episodic and their impact is sufficiently constrained. Wilderness does, however, exclude treatment as a *resource* for other, more intrusive, sorts of use. As an illustration of the point, the Cockell and Horneck proposal for planetary parks on Mars set out the following restrictions [11] (p.294):

1. no spacecraft/vehicle parts to be left within the park;
2. no landing of unmanned spacecraft within the park;
3. no waste to be left within the park;
4. access only on foot or surface vehicle along predefined routes or landing by rocket vehicle in predefined landing areas;
5. all suits, vehicles and other machines used in the park to be sterilized on their external surfaces to prevent microbial shedding

We might raise questions about some of these restrictions, and we might want to add others. For example, the landing of unmanned spacecraft would surely be less polluting than the landing of manned craft. However, these proposals do look like the kind of rules that might be applied to protect localized regions of Mars, such as portions of the Valles Marineris or Olympus Mons, as wilderness sites. In line with this, we will accept that any plausible wilderness approach will be *locally* restrictive in particular places. But it does not follow from this that such an approach must be restrictive or excessively constraining *overall*, with respect to a planetary surface or any other location. Indeed, the Cockell and Horneck approach is specifically focused upon the local protection of special and exemplary sites and might not on its own be sufficient if we support the idea that (for whatever reason) a planet is due global environmental protection.

As a final clarification, by proposing that seven-eighths of the Solar System ought to be left as wilderness, the suggestion is not that seven-eighths of every individual object ought to be reserved. Such an approach would be locally demanding in an excessive way. Rather, the suggestion is that seven-eighths of

---

[8] This will also leave the issue of the standing of 'restored' environments an open one.

*everything usable* within the Solar System ought to be reserved. It is not a claim about any specific object. The reserving of some places and some things is, after all, likely to be more important than the reserving of others. Having to make decisions to reserve *a* rather than *b* is sometimes unavoidable and often desirable. Various narratives about ethics and the pragmatics of policy and stakeholdership may be drawn upon in order to help make such decisions. Different narratives may point in different directions and raise competing local concerns, yet they may do so within a broader agreement about how much *overall* should be brought into use.

## 3. EXPONENTIAL GROWTH AND DOUBLING TIME

Estimating the size of the future space economy is an exercise in understanding exponential growth. Such growth is a familiar terrestrial phenomenon. Biological systems grow exponentially unless constrained by factors such as disease, resource availability or outside influence. The growth of the rabbit population in Australia is a familiar example. Two dozen rabbits released in 1859, to provide sport for hunters, increased within 6 years to 22 million rabbits[9]. Economic growth exhibits similar features albeit on a more modest scale. Compound interest is a famous case. A single cent placed on the first square of a chess board with the running sum then successively doubled on each square will yield $9,223,372,036,854,780,000 by the final square, only 63 steps away. A more than reasonable return for a low initial investment.

Growth rates for GDP are also exponential– the economy grows by some fraction of its previous value each year. The major economies of the world are, in part, the result of such exponential growth with seemingly modest annual increases generating extremely large results over time. Table 1 shows how annual percentage growth rates translate into cumulative growth factors over timescales of up to a century. Towards the high end of the scale, China's economy has grown at almost 10% a year over the past 30 years[10]. Over the course of a century this would produce a cumulative growth factor of nearly 14,000 relative to the starting figure. Historical growth rates for the US have, by comparison, been much lower, averaging out at around 2% a year for the past century. Yet, this has still been enough to allow it to play a sustained role as the world's dominant economic power[11]. European GDP figures are comparable to those for the U.S. Exponential growth allows modest initial numbers to become large over a surprisingly small number of cycles. Beyond a certain point, growth gathers momentum which is initially hard to imagine and then hard to contain.

---

[9] http://www.petefalzone.com/handouts/exp-growth-rabbits-australia.pdf
[10] http://www.indexmundi.com/china/gdp_real_growth_rate.html
[11] 1870-2001, http://socialdemocracy21stcentury.blogspot.com/2012/09/us-real-per-capita-gdp-from-18702001.html

Table 1 makes it easy to see why at least some space enthusiasts believe that today's $300B Space Economy[12] will quickly outpace the strictly-terrestrial economy, given that the latter is limited by supply and ecology as well as political constraints. Space may seem to be freer of such considerations. In this vision of rapid transformation, China-like 10% compounded growth would propel the space economy to a $4200 Trillion economic value within a century. By comparison, World GDP in 2012 was only $71.6 Trillion, almost 60 times smaller[13]. At anything close to a sustained 10% growth rate, the space economy would, indeed, quickly come to dominate.

Table 1 Cumulative growth factors at plausible growth rates.

| Growth Rate | Doubling time (years) | 25 years | 50 years | 75 years | 100 years | 200 years |
|---|---|---|---|---|---|---|
| 2% | 35 | 1.6 | 2.7 | 4.4 | 7.2 | 52 |
| 3% | 23.5 | 2.1 | 4.4 | 9.2 | 19.2 | 370 |
| 3.5% | 20 | 2.4 | 5.6 | 13 | 31 | 973 |
| 5% | 14 | 3.4 | 12 | 39 | 130 | 17,300 |
| 8% | 9 | 6.9 | 47 | 321 | 2199 | 4.8 million |
| 10% | 7.3 | 10.8 | 117 | 1270 | 13780 | 190 million |

However, exponential growth, especially exponential economic growth, ordinarily tapers off at some point. We may not expect the space economy to consistently sustain a 10% growth rate even if it reaches that rate (or even higher) during its early stages. Beyond a certain stage in its development, the accumulation of new resources simply becomes more difficult. Increasingly marginal opportunities have to be considered. Internal political pressures, variations in the demand for exports, and exhaustion of the sources of competitive advantage may also act as a partial break. We need only look at the Chinese economy to see this, with three decades of growth slowing considerably after 2015, falling back from 10% to around 6.5%. Whether or not a system slowdown occurs early enough to avoid major growth-related problems is, however, a contingent matter. There may be various slow-down mechanisms, but it is not obvious that there is any such mechanism that will always or necessarily kick-in early enough, or hard enough, to prevent major difficulties from emerging.

---

[12] The Space Report, 2016 gives a figure of $322.94 bn.
[13] http://databank.worldbank.org/data/download/GDP.pdf

Gauging the pace of exponential growth is, therefore, important. The timescale that is ordinarily used is the "doubling time," i.e. the time that it takes to double in size. Even a modest succession of such doubling times illustrates the potential for the rapid emergence of growth problems. Over only three doubling times, an eighth becomes successively a quarter, then a half and then 100%. For the Australian rabbits, we know that the doubling time was 3.6 months, since it took 6 years for the initial 24 rabbits to reach a million times larger number. Remarkably, this is only 20 doublings. If we re-wind by just 3 doubling times there would have been only one-eighth as many rabbits, which is a lot, but not yet an ecological problem of a comparable scale. Yet, it would be a warning point, an indication that growth was out of control, and likely to reach a crisis.

This is the mathematics behind why we advocate setting a threshold of 3 doubling times before a state of super-exploitation. Reaching one-eighth would tell us that we were uncomfortably close to exploiting the entire usable resources of our solar system. On the assumption that growth rates will be nearer to 3% than to 10%, this would leave only a half century for an economic transition to a stable state (requiring no further resource utilization) to be completed. A difficult predicament best avoided, it would be safest if the transition were already complete before super-exploitation. Expressed as a further general principle: we ought not to deliberately expand beyond the point at which a future generation of humans could (reliably and safely) carry out an emergency slow down.

What might then, afterward, be permissible under conditions *without* exponential growth is beyond the scope of the present paper. Ideally, given that we do have conditions of exponential growth, we should try to avoid tripping over the one-eighth mark, and should certainly be thinking about how to slow down well beforehand. We accept that reasonable people may disagree on the precise tripwire value to set, and how much stopping distance is necessary, but it will surely be *at least* 2 doubling times (given familiar levels of economic growth).

Why not go back farther than 3 doubling times? Table 1 shows that small errors in estimating the growth rate grow into large discrepancies in the predictions by 5 doubling times, which could lead to charges of crying wolf. (C.f. the 50% different predictions for a ½% difference in growth rate, from 3% to 3.5% at 100 years.) Errors in estimate growth rates can work in both directions (leading to either an underestimation or an overestimation of the time left to act. Under conditions of imprecision about such matters, our choice of an estimated 3 doubling times seems prudent without being excessively demanding.

The issue of exactly how demanding it would be to adopt a one-eighth principle is tackled below in section 5. However, the practicality of the principle will also depend upon arriving at some viable way to measure overall resource usage, a problem tackled in section 4.

**4. A SECTORAL APPROACH TO MEASUREMENT**

To form a more precise idea of what a precautionary one-eighth resource exploitation limit would mean in practice we have to ask questions of a more fine-grained flavor: whether the one-eighth should be measured by mass, volume, surface area or number. Let us call this the *measurement problem*. Some objects or places may, after all, be suitable candidates for protection from development because of special scientific interest, or on ethical grounds related to their uniqueness, but more so with regard to certain of their characteristics rather than others. In some locations, surface alteration may be regarded as more significant than sub-surface alteration. The Earth itself is an obvious example. Surface or near surface protection has always been the priority of ecologists and geologists. Even on James Lovelock's controversial *Gaia* hypothesis, the terrestrial system that he is seeking to protect reaches downwards only for 100 miles (2.5% of the Earth's radius) and no further [12] (p.19).

While there may sometimes be reasons for protection of the deep interior of a planet, these are likely to be reasons of a different sort from those connected to the portions of a planet that we (or our near descendants) might interact with in a far more direct way. In line with this, our assumption is that surface area is more likely to be a suitable unit for measurement on planets and large moons while mass is more likely to be a suitable unit of measurement in relation to asteroids. The obvious exceptions to this are Ceres, Vesta, Pallas and Hygeia, the four largest asteroids. Their scale might lead us to prioritize their surfaces, or else it might lead to their inclusion within the seven-eighths of the mass of the asteroid belt that should be left as wilderness. This may, however, be locally demanding given that the mass of the asteroid belt is spread out over a large volume and their mass accounts for around half of the total belt mass. Logistics alone will then make them prime targets for the extraction of materials., and possibly also as bases of operations. Protection and ease of access may sometimes point in different directions.

Prioritizing different measures - surface area for planets and large moons[14], and mass for the asteroid belt (with the noted exceptions) - poses something of a dilemma about how we can aggregate our figures and arrive at an overall assessment of how close we are to the one-eighth limit at any given time. However, this is less of a problem than it initially appears to be. The point of the exercise is crisis avoidance, *not* absolute precision of measurement. The aim is to establish a suitable prudential tripwire to avoid the harms and injustices of super-exploitation, and not quantification for some other reason.

Bearing this in mind, we may simply go 'sectoral.' With regard to a particular sector or sub-sector (i.e. a specified area, location or object) we might have reasons to use one measure rather than another, e.g. mass rather than surface area. But whatever local value is in use, we should take one-eighth maximum usage as the sector default. In line with what has already been claimed, defaults are not absolute restrictions. 25% usage *however measured* in one sector or sub-sector, will aggregate to 50% overall usage when it is considered alongside another *comparable* sector where usage is running at 75% (however measured). Criteria for comparability could include scale, rarity or accessibility. No doubt there are other suitable criteria which might reasonably license us to say that one sector and another can be considered together. This will allow for sectoral trade-offs to be made with heavier utilization occurring in some places than in others. In so doing, protection is made more practical. Systems of double or multiple weighting might even be brought into play in order to encourage development *away* from particular areas. Mass extracted from Ceres, for example, might count double compared to mass extracted from smaller asteroid belt objects. This would operate as an incentive to look elsewhere. Moves of this sort should, however, lead us to consider how much has been used by several different units of measurement. It would, no doubt, be important to have ways to gauge the impact of pragmatic trade-offs to ensure that the underlying principle was being respected.

---

[14] There are 6 solar system moons larger than our Moon (1738 km dia.): 3 of the 4 Galilean moons of Jupiter: Io (1810 km), Ganymede (2600 km) and Callisto (2360km), but not Europa (1480 km); Saturn's moon Titan (2440 km); Neptune's moon Triton (1900 km) [11]. Pluto's largest moon, Charon, is a little smaller (1212 km) than our Moon.

There will, and should, be arguments about how best to aggregate the data and about what makes one sector or cluster of sectors comparable to another or a good candidate for special weighting. Disagreement about such matters would not imply the absence of a useful answer, or even that some answers are better than others. The economics of space are unlikely to yield sudden universal agreement when arguments about terrestrial economics, and the aggregation of data, have failed to do so. Our case requires only that there is *at least* one viable way to aggregate the data. We expect that there will turn out to be several, but the one proposed looks like the most obvious, as well as being flexible in a way that may help to answer the charge that the principle will be too restrictive.

There are, of course, various downsides to this (and perhaps to any) approach. Local variations in the way that the *measurement problem* is dealt with, will create possibilities for a skewed pattern of protection in the light of preferential demand for particular resources. If a resource is concentrated near to the surface (in the way that Helium 3 is on planets, moons and asteroids) there will be pressure to use mass as measure, in order to leave the way open for a greater extent of mining. Skewing can, perhaps *will*, lead to some inappropriate local solutions. We are not, however, proposing a system to avoid the usual difficulties that *any* set of shared norms or regulatory structures are likely to face. Rather, our assumptions are that at least some level of systemic imperfection is unavoidable, but interest-driven skewing of measurement would not always be worth tackling at the expense of an over-centralization of authority.

Indeed, it is our view that an appropriate response to the *measurement problem* should prioritize the demands of ongoing processes of negotiation and trade-offs between multiple legitimate stake-holders rather than those of a unitary and centralized command control system. It is far from obvious that, in the long run, any system of the latter sort would be able to overcome the difficulties of scale involved in exercising authority over the entirety of the solar system. (Less ambitious command systems might be viable.) Effective approaches towards wilderness protection in space, as a whole, are more likely to benefit from a pragmatic attitude towards multiple (sometimes competing) interests and towards shifting political trends. Wilderness protection in space cannot, therefore, simply be opposed to economic development in the way that has sometimes been effective on the Earth. The survival of large tracts of ancient European woodland, the *Belovezhskaya pushcha*, through the turmoil and industrial development of the 20th century, may reasonably be credited to Soviet era command control, and the consequent ability to over-ride any competing interests.[15] This is not, however, a model that could be of more than local significance in space and time. It has not actually outlasted the Soviet era. Much of this woodland is now at risk.

Reinforcing this pragmatism about economic and political matters, the overall focus of the one-eighth principle has the further advantage that it is consistent with pragmatism about the importance of particular objects. Sufficiently weighty considerations would, in all likelihood, trump the case for protection of *any* particular object or place. However, the considerations might have to be fictionally extreme before we would be prepared to sacrifice the only known remnant of extraterrestrial life, for example, in order to extract some valued resource. This is, however, a thought experiment and not a problem that we currently face. Additionally, what we consider important may not be a priority for future generations. An approach towards wilderness that starts from the absolutely non-negotiable standing of *any* particular cluster of objects or places may not be a persuasive consideration across generations. It is a strength of the one-eighth principle that it is flexible when it comes to the question of exactly *what* is brought into use at any given time. The principle requires only that we "tension" decisions: when we exceed the quota in one respect, and in any given sector, some *appropriate* compensation in the form of additional restraint then has to be made elsewhere. We do not get something for nothing. In this way the principle helps incorporate the externalities of unfettered expansion into planning.

Admittedly, there will, and perhaps should, be some skepticism about any proposal for a trade-off system in the light of our experience of carbon-emissions trading. Most notoriously, the Wikileaks exposure of cases in which greenhouse gases have themselves been commodified, cases in which they have been produced outside of the industrialized West for the specific purpose of being traded off against emissions in the US and Europe.[16] However, known problems with trade-off systems are perhaps to be expected because of the scale of the difficulties that they seek to address. Such difficulties might equally be taken as guidance for the kind of future trade-offs that are most likely to work, and those most likely to fail. There is, after all, a significant difference between systems operating in an imperfect way and systems being unfit for purpose. We have, throughout, assumed system imperfection and pragmatism about how best to respond to it.

---

[15] In line with UNESCO practice, we have used the Russian name from before the dissolution of the Soviet Union, although the woodland spans several countries.
[16] https://www.scientificamerican.com/article/carbon-credits-system-tarnished-wikileaks/

### 4.1 The Need for Scaled Exploration of the Solar System

A practical upshot of the proposed approach is that we need to inventory the resources of the Solar System carefully and at a sufficiently early point in time in order to know just what lies out there. The argument underpins the need for certain kinds of science, and not simply because of its convenient spin-offs. Additionally, we need further clarification of the varying ethical grounds for protection before we can advance provisional *overall* proposals for which parts of the Solar System we should use as resource, and which parts to reserve. We need the niches for key contending ethical theories to be filled and the reasons for protecting one place rather than another to be clearly stated.

Any viable *overall* protection proposal would involve a good deal of detailed ethical argumentation and a vastly enhanced program of visits first to the inner, and eventually to the entire, Solar System. (These should be robotic to reduce any initial impact.) Worldwide, the present rate of planetary mission launches is 15/decade[17], with about 60% being NASA-led. At this rate, even just the nearly 200 worlds of the solar system that gravity has made spherical would take 130 years to visit once. This is a significant fraction of the 400 years we may have. For now, we could ignore the outer solar system, beyond the Main Belt. Even then, the number of bodies to investigate is still large, most of them being asteroids. Below some size a statistical approach will have to be adopted.

An overall survey of ethical, economic, social and policy considerations as well as a comparable survey of material resources, would require a team of researchers from multiple disciplines, and is therefore beyond the scope of the present paper. A rationale for undertaking such an overall survey is, however, built into the one-eighth principle. Deliberations over such considerations should begin to decide how to delineate which eighth is best chosen for exploitation.

### 5. HOW RESTRICTIVE IS THE ONE-EIGHTH PRINCIPLE? IRON RESOURCES IN THE ASTEROID MAIN BELT

We are required, as a point of social ethics, to accept reasonable constraints upon our self-interest in order to meet basic standards of justice between one another and (arguably) between ourselves and near future generations. This is a precondition of having any sort of stable and lasting human society. However, we will take it that a *livable* ethic for society at large cannot ask for too much. More precisely, a reasonable social ethic cannot ask for anything so demanding that it is impossible, inconsistent with what we know about human psychology, or otherwise so demanding that it belongs only in the domain of private sacrificial commitment of a sort associated with political and religious ideals. The one-eighth restriction may seem to fall foul of this constraint. It may seem to ask for too great and prolonged a forgoing of opportunities. In this section, we will try to show that this is not the case.

---

[17] J. McDowell, 2017 private communication, based on data available at planet4589.org. See also: https://en.wikipedia.org/wiki/List_of_Solar_System_probes.

Adoption of the principle would leave a large scope for economic development and for the advantages that it offers.

To make this case, we will focus upon the mining of the asteroid Main Belt. This is where most of the raw materials that we are likely to appropriate in space are to be found. Setting aside the special cases of the largest objects in the belt (Ceres, Vesta, Pallas, Hygiea and half a dozen other objects of similar order to the latter) the vast majority of objects in the belt do not display anything like the uniqueness of planets or moons. Concerns about their integrity (in the sense relevant to the ethics of protection) do not apply. We have no special reason to protect surface area rather than mass, which is a more useful measure for mining operations. But just how demanding would the one-eighth principle be in this pivotal case, when applied to the mass of the asteroid belt? It would certainly require us to leave a great deal behind, unused. But it would also set an upper limit upon use that could not easily be regarded as excessively restrictive. More precisely, one eighth of the iron ore in the asteroid belt would still be more than a million times greater than all the known iron reserves on Earth.

Backing up this claim requires us to reverse a familiar calculation. In the past, meteor composition has been used to help estimate the composition of the Earth down to its core. If we run this in reverse, using a best estimate of the fraction of the Earth that is iron, it will give us a guide to the overall composition of the asteroid belt. The Earth has an estimated iron mass fraction whose limits sit just under and over 30 wt%. (This is 'bulk Earth,' and not just its Fe-Ni core [14, Table 6] "Limits on the composition of the core and bulk Earth.") Given that the millions of asteroids within the belt have a total estimated mass of ½% of the Earth's mass[18] or ~3x10$^9$ bn.mt [15], then, if the Belt has the same overall composition as the bulk Earth (i.e. 30% iron) then the Belt will contain around 10$^9$ bn.mt of iron. However, most of the Earth's reserves are not actually extractable (we can hardly hollow-out the planet's core). In contrast, the metals in the asteroid belt are much more accessible as the cores of the original bodies were broken up in collisions, exposing them as the metallic asteroids. Even leaving aside the 60% of metals that lie in the four largest asteroids [16], where this collision process was not effective, 1/8 of the total mass of accessible asteroid iron is very large.

What we need for comparison between what we can use *here* and what we might use *out there* is an estimate of only those terrestrial iron reserves that could, in principle, be extracted for use. As of January 2017, the U.S. Geological Survey estimate of such reserves is a far more modest 82 bn. mt.[19] Even one eighth of the asteroid belt iron then would be more than a million times this amount.

Another way to put matters would be to ask: "Could we build a Dyson sphere [17] with that much iron?" More modestly, we might begin with a ringworld. A ringworld is a continuous circle constructed around the Sun at about the distance of the Earth's orbit, so that water is liquid on its inner face [18]. The first step would be putting a single ring made of iron girders around the Earth's orbit. (The structure needs to be strong as it will be rotating sufficiently to produce Earth gravity on its inner surface through centrifugal force.) We estimate that even with the restriction imposed by the one-eighth principle, and setting aside the four larger asteroids, we could still build 2 million Earth-orbit-girdling rings from Main Belt iron.[20] That should be enough to go on with, though a Dyson sphere may be out of reach. Of course, the suggestion here is not that we actually do this. The example is, rather, more of a visualizing aid. (An actual Ringworld would require some major solar system engineering.)

[18] 1 Earth mass = 6x10$^{12}$ billion metric tons (bn.mt) [3].

[19] https://minerals.usgs.gov/minerals/pubs/commodity/iron_ore/mcs-2017-feore.pdf

At some point in the future iron for 2 million Earth-orbit-girdling rings might still become a restrictive quantity. Especially if we imagine multiple bustling human worlds. However, the solar system has only a limited number of planets and moons that are ever likely to be suitable for a stable human presence.  Even adding some off-world O'Neil-like habitats [19], matters would not change until we start imagining a truly vast number of such habitats. Short of such a scenario, a restriction to one-eighth of the asteroid belt iron hardly seems likely to be any sort of overall demanding constraint. Indeed, it would allow for such a large scale of expansion that we might then wonder whether or not we need a one-eighth principle at all because one eighth is unlikely to be reached.

A circular economy can help. However, recycling can never be 100% efficient. All the material used as propellant to move rockets around the solar system is certainly lost. A stronger constraint could be that much of the material mined will have gone into making habitats for humans, especially so once off-world O'Neil habitats become the norm. Those people will object to having their homes re-cycled, especially when those habitats are providing the air they breathe.

However, such a limit performs a useful task. Exponential growth removes the room for complacency in the face of the apparent security that vast solar system resources seem to offer. We may, instead, wonder whether the million times more plentiful resource in the asteroid belt is really going to be such a vast amount once our tendencies to expand and to consume are taken into account. As we saw in the case of the Australian rabbits, 1 million is very close to only 20 doublings. [$2^{20}$ = 1,048,576.] 1 million times the Earth's current iron reserves will leave less room for doubling than we might imagine. The 17 doublings limit of the 1/8 principle might still leave time for adaptation if the doubling time for economic expansion is quite long. We might then be tempted to defer consideration of the problem until it becomes pressing.

---

[20] The circumference of the Earth's orbit is $2\pi \times 1AU = 2\pi \times (1.5 \times 10^8$ km$) = 9.42 \times 10^8$ km. I-beams use for tall building construction have masses of 38.1-57.7 mt/km for (https://www.alibaba.com/product-detail/carbon-structural-steel-h-beam-mild_60425467973.html). Using 50 mt/km, the mass of iron needed to girdle the circumference of the Earth's orbit is $5 \times 10^{10}$ mt. We estimated that the Main Belt has $1 \times 10^{18}$ mt of iron; with our constraints that reduces to $1 \times 10^{17}$ mt,  or $2 \times 10^6$ 'girdles'.

But is the doubling time likely to be so favorably long? It is, of course, difficult to tell. What we have to go on are analogies with doubling times in the past. Since the early industrial revolution [20, p.178] estimates that we have moved from global production of around half a million tons of pig iron around 1800 to half a billion tons of steel produced in 1994, i.e. a factor of a thousand increase. The 200 years separating then from now would then be equivalent to 10 doubling times of 20 years. USGS data on iron ore production suggests a broadly similar doubling time, given world production of 1 billion tons of crude ore in 1994 and 2.2 bn. tons in 2016, only 22 years at a fairly modest 3.5% growth rate.

If the space economy matched that a 3.5% growth rate, then 400 years from now the ratio between asteroid reserves and annual production would be the same as that which exists now between terrestrial iron production and terrestrial iron reserves. We would, at that point have only 60 years (i.e. three doubling times) before exhaustion. The exhaustion involved in reaching a point of super-exploitation would, however, be even more serious than exhaustion of untapped Earth iron, given that we would have no larger body of accessible metals to which we could then look without venturing beyond the bounds of the solar system itself.

The outer solar system that lies beyond the orbit of Neptune and is roughly 10 times more distant than the asteroid Main Belt, contains the bodies of the Kuiper Belt. Though little explored as yet, these bodies may contain as much as 10 times the resources of the Main Belt, about an 1/10 an Earth mass of material [21]. Does this change our argument? Yes, but not by much. The ravenous nature of exponential growth means that the Kuiper Belt buys us little more than 3 more doubling times, i.e. another 60 or so years. After that there is the Oort cloud of comets, which is at least 100 times further away even than the Kuiper Belt [22] and contains, at a very rough estimate, an Earth mass of material [23], so $2x10^{12}$ billion mt of iron, about 200 times that of the Main Belt asteroids and about 20 times that of the Kuiper Belt. If we can exploit the Oort cloud too then we gain another 4 doubling times, or some 80 years (at 3.5% growth).

Then we are done. It is another 20 times further from the Oort cloud to Proxima Cen, the nearest star. Going half way to Proxima Cen would be enough to find another 4 Earth masses in interstellar asteroids [24]. However, journey times, which are already measured in decades for the Oort Cloud, increase to centuries for anything but microscopic masses at these interstellar distances[21].

If new physics and technology fail to emerge, then all we have to rely on for additional material are interstellar asteroids arriving from beyond the solar system. It is already possible to build a mission that could rendezvous with a duplicate of the first interstellar asteroid [25] *'Oumuamua* [26]. Thanks to the discovery of *'Oumuamua*, we now have an estimate of the rate at which they enter the solar system [7]. That rate is about 1 per year [25] bringing about a million tons per year to us. At the advanced stage of the solar system economy that is a tiny supply, only equivalent to our usage in the early 19th century. While these numbers could be off by a factor 10, they demonstrate that the solar system is effectively a closed system.

---

[21] See Breakthrough Starshot Project, https://breakthroughinitiatives.org/initiative/3

## 7. CONCLUSION

To summarize, while we remain dependent upon the resources present inside the Solar System, and while economic growth remains exponential, we should regard, at most, one-eighth of the solar system as humanities to use. The remaining seven-eighths of the solar system should be left as space wilderness. (In the thin sense that it should not be brought into regular economic use as a resource.) Failure to do so will mean that future generations will have insufficient 'breaking distance' after only a few centuries of exponentially growing economic activity/resource utilization. If unchecked, such growth will tend towards a point of super-exploitation, i.e. a situation of resource depletion where new resources cannot readily be brought into use, even in an emergency situation. The dangers of super-exploitation, for a space-faring civilization whose limits are set by the bounds of a single solar system, are too great to be set aside.

On a timescale of less than a millennium we could have super-exploitation of the entire solar system out to its most distant edges. A millennium is a long time to look forward, but is not long in human history, and is tiny in solar system history. Facing up to this inevitable consequence of economic growth demonstrates that a circular economy with near-perfect recycling of raw materials is simply a requirement. If we can begin on a mere planetary scale now, we will be prepared to adapt to a solar system scale as we gain that capability.

While the principle that we should use, at most, one eighth of the resources present in the solar system relies upon an ethical duty towards humanity (or, more specifically, a duty to near human generations) it does not rest upon any particular and perhaps idiosyncratic ethical theory. Rather, it rests only upon a minimal set of ethical claims which happen to be common to a variety of different theories. Rights based approaches, consequentialist approaches and virtue ethical approaches all tend to accept that we do indeed have a duty to consider the interests of future generations. We have, throughout, sought to keep the ethical claims minimal.

There are, finally, at least two practical consequences. First, a full science-based inventory of the solar system should be undertaken in order to apply the one eighth principle wisely. This implies a great increase in the rate of exploration if it is to be completed in a small fraction, say 10%, of the time available before the earliest likely point of exhaustion, i.e. about 40 years. Second, there is a need for more detailed ethical theories about the various reasons for protection and use, as well as deliberations about wise policy for choosing which eighth should be exploited.

## Acknowledgements

ME thanks his daughter, Camilla J.F. Elvis, for insisting that he think about these issues. ME also thanks Jonathan McDowell for the planetary mission launch statistics, and the Aspen Center for Physics, which is supported by National Science Foundation grant PHY-1607611, for their hospitality while this paper was first drafted, and later completed. This research did not receive any specific grant from funding agencies in the public, commercial, or not-for-profit sectors.